\documentstyle[hip-artc]{article}  

\volnumber{5}  \edyear{1997}  \frompage{000} \topage{000}                
\recrevdate{1 January 1996; revised version 1 January 1997} 			 
\title{Properties of superdeformed fission isomers in the cranked
relativistic Hartree-Bogoliubov theory.} 
\authors{
{\twerm A.\ V.\ Afanasjev$^{1,a}$ and P.\ Ring$^{2}$ %
}\\[2.812mm]
{\normalsize
\hspace*{-8pt}$^1$ Physik-Department der Technischen Universit{\"a}t
M{\"u}nchen,\\ D-85747 Garching, Germany\\[0.2ex] 
}}
 
\abstract{The rotational and deformation properties of
superdeformed fission isomers in the $A\sim 240$ mass 
region have been investigated within the framework of 
the cranked relativistic Hartree-Bogoliubov theory.
The dependence of the results of the calculations on
the parametrization of the RMF Lagrangian has been
studied. The rotational properties are best 
described by the NL1 force.}
 
\begin{document}
 
\maketitle

  The region of $A\sim 240$ is the first one where the 
superdeformed (SD) shapes have been discovered experimentally 
\cite{Pol.62} in the fission isomers in 1962. According 
to the measurements of the moments of inertia and the 
quadrupole moments they are shape isomers with the deformation 
much larger than the usual ground state deformation. Despite 
the fact that during almost 40 years numerous experimental 
efforts have been made for the study of fission isomers, the 
experimental data on their properties in some respects is 
more limited than the one collected in the last 15 years in 
the regions of superdeformation at high spin such as 
$A \sim 60, 80, 130,
150$ and 190 \cite{BFS.96}. Recent advances in the experimental 
techniques leading to the observation of hyperdeformed (third) 
minimum \cite{U236HD,U234HD1,U234HD2} and vibrational structures 
in the SD (second) minimum \cite{Pu240a,Pu240b,Habs-priv} 
revived the interest to this mass region of superdeformation. 

   A number of theoretical investigations has been performed 
in this mass region studying different properties of fission
isomers within the frameworks of the phenomenological macroscopic 
+ microscopic method and non-relativistic microscopic theories 
based on the zero-range Skyrme forces and finite range Gogny 
forces, see Refs. \cite{HO.75,DNF.84,CNSPJ.94,BGG.89,KBFHW.94} 
and references therein.   Recently, some properties of fission 
isomers in $^{226}$Ra, $^{232}$Th and $^{240}$Pu have been 
studied in the relativistic mean field (RMF) theory by the Frankfurt 
group \cite{RMRG.95,BRRM.99}. These investigations, however, have 
been performed in the BCS approximation using a schematic treatment 
of pairing in the constant gap approximation with the pairing 
parameters adjusted to the experimental data in the first 
minimum.

 The present investigation is aimed on a more systematic study 
of the properties of SD fission isomers within the framework 
of RMF theory in the present state of the art. The results of 
this study will be presented in a forthcoming article 
\cite{AR.00}. In the present contribution, we will
concentrate on the even-even nuclei in which rotational
structures have been observed so far \cite{H.89}, namely 
on $^{236,238}$U and $^{240}$Pu nuclei.

\vspace*{-18pt}
\begin{table}[h] 
\caption[]{Kinematic moments of inertia ($J^{(1)}$), charge 
quadrupole ($Q_0$) and mass hexadecupole ($Q_{40}$) moments 
of SD band in $^{240}$Pu calculated at rotational frequency 
$\Omega_x=0.01$ MeV with different parametrizations of the 
RMF Lagrangian. In the case of the NL1 force,  
the results of the CRHB calculations with and without 
APNP(LN) [marked as 'LN' and 'Unpr'] and the results of 
the CRMF calculations with no pairing [marked as 'CRMF'] 
are given. For other forces, only the results of the CRHB 
calculations with APNP(LN) are shown.}
\begin{center}
\begin{minipage}{0.9\textwidth} 
\renewcommand{\footnoterule}{\kern -3pt} 
\begin{tabular}{cccccc}
\hline\\[-10pt] 		
Quantity               & NLSH & NL3 & NL1               &  NL-Z & Exp. \\ 
         &      &     & (LN/Unpr/CRMF)  &  \\
\hline\\[-10pt]			
$J^{(1)}$ [MeV$^{-1}$]  & 138.2 & 144.4 & 155.2~/~177.2~/~189.1 & 157.3 & 149.3 \\ 
$Q_0$ [$e$b]              & 33.0 & 35.5 & 38.2~/~39.6/~39.5   & 39.7  &  \\ 
$Q_{40}$ [fm$^4$]               & 36121 & 41901 & 49425~/~56329~/~55533   & 53328  &  \\ 
\hline 
\end{tabular}
\end{minipage}
\renewcommand{\footnoterule}{\kern-3pt \hrule width .4\columnwidth 
\kern 2.6pt} 			
\end{center}
\end{table}
\vspace*{-8pt}

  As a theoretical tool we are using the recently developed 
Cranked Relativistic Hartree-Bogoliubov (CRHB) theory \cite{CRHB} which has been 
very successful in the description of the properties of SD 
bands in the $A\sim 190$ mass region \cite{A190,CRHB} and 
rare-earth nuclei \cite{J1Rare}. Compared with previous 
relativistic studies in the $A\sim 240$ mass region, it 
has the following advantages: 
{\bf (i)} the cranked RMF equations are solved on the 
Hartree-Bogoliubov level, {\bf (ii)} the finite range 
D1S Gogny force is used in the particle-particle channel, 
{\bf (iii)} approximate particle number projection is 
performed by means of the Lipkin-Nogami method (further
APNP(LN)), {\bf (iv)} the cranking model approximation is 
employed which allows to study the rotational properties 
of fission isomers and {\bf (v)} this theory is formulated 
in three-dimensional Cartesian coordinates which allows to 
study the possible appearance of triaxial deformation. 
Since this study is concerned with the properties of 
the states in the second minimum, only reflection symmetric
shapes are considered. This approximation is justified
since octupole deformation becomes important only after 
the second well (see for example Refs.\ 
\cite{CNSPJ.94,RMRG.95}).

   The calculated moments of inertia of the SD isomers in $^{236}$U 
and $^{240}$Pu are shown in Table 1 and Fig.\ 1. The change of the 
force from NLSH \cite{NLSH} via NL3 \cite{NL3} and NL1  \cite{NL1} 
to NL-Z \cite{NLZ} leads to the increase of the kinematic moments of 
inertia, charge quadrupole and mass hexadecupole moments (Table 1). 
The experimental moment of inertia of the SD band in $^{240}$Pu \cite{H.89} 
is located between the results of the calculations with NL3 
and NL1 (see Table 1). The use of the NLSH and NL-Z forces 
for the RMF Lagrangian leads to larger deviations from experiment. 
The results of the calculations with NL3 underestimate the
experimental moments of inertia by $3-5$\%. This is also the case 
for $^{238}$U where experimental and calculated (with NL3) moments 
of inertia are 153.1 and 142.8 MeV$^{-1}$. The difference between 
$J^{(1)}$ moments of inertia of $^{236,238}$U nuclei is 3.8 MeV$^{-1}$ 
in experiment while it is only 0.6 MeV$^{-1}$ in the calculations with 
NL3. The NL1 force describes the absolute values of the moments of 
inertia and their difference in $^{236,238}$U better than NL3. In 
these nuclei, the $J^{(1)}$ values calculated at $\Omega_x=0.01$ MeV 
are 150.5 and 154.7 MeV$^{-1}$ which agree very well with the
experimental values of 149.25 and 153.06 MeV$^{-1}$. On the other 
hand, the calculations with NL1 somewhat (by $\sim 4$\%) overestimate 
the experimental moments of inertia in $^{240}$Pu (see Fig.\ 1 and
Table 1).

\begin{figure}[htb]
\vspace*{0.0cm}
\epsfxsize 11.5cm 		
\epsfbox{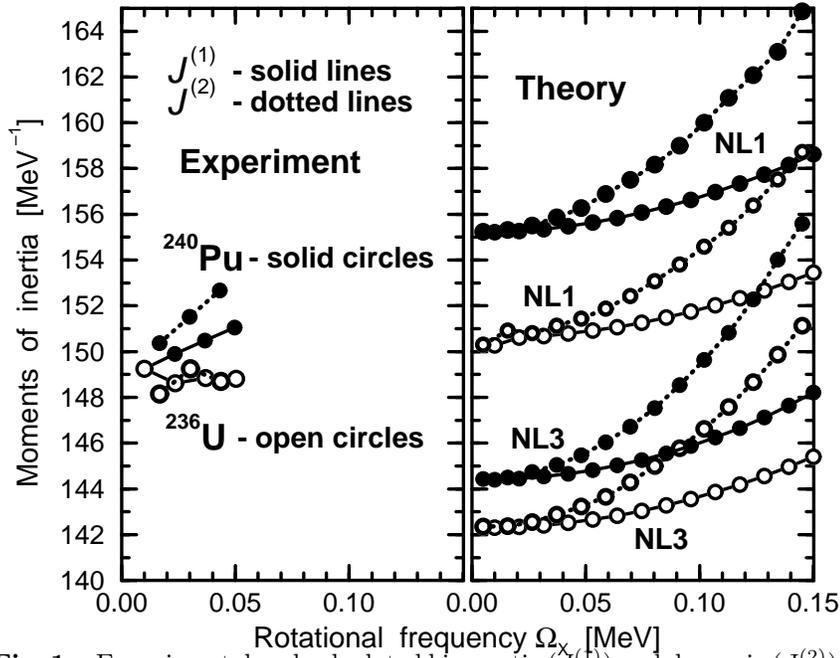}
\vspace*{-0.8cm}
\caption[]{ Experimental and calculated kinematic $(J^{(1)})$ and 
dynamic $(J^{(2)})$ moments of inertia of SD fission isomers in 
$^{236}$U and $^{240}$Pu. The notation of the lines and symbols 
is given in the figure.}
\label{fig-j2j1}
\end{figure}
\vspace{-0.4cm}

  The CRHB calculations indicate that kinematic and dynamic
moments of inertia increase with increasing rotational 
frequency $\Omega_x$ (Fig.\ 1). In addition, the difference 
between these moments grows with the increase of 
$\Omega_x$. The experimental data in $^{240}$Pu shows
such features, while they are not seen in $^{236}$U. These 
features are predominantly due to the gradual alignment of 
the $N=8$ neutrons and $N=7$ protons and a smooth decrease 
of pairing correlations with increasing $\Omega_x$. They are 
similar to the ones observed in the $A\sim 190$ region of 
superdeformation, see Refs.\ \cite{A190,CRHB} and references 
quoted therein.
   
   The importance of APNP(LN) for the description of the 
rotational and deformation properties of SD fission isomers 
is clearly seen on the example of the calculations performed with 
the NL1 force (Table 1). The CRMF calculations \cite{AKR.96} with 
no pairing give $J^{(1)}=189.1$ MeV$^{-1}$ which is below 
the rigid body moment of inertia ($J_{rig}=211.7$ MeV$^{-1}$) 
defined from the density distribution. The inclusion of 
pairing somewhat decreases the calculated moment of inertia 
which still considerably exceeds the experimental value. 
Only the CRHB results with APNP(LN) come close to 
experiment. One should also note that in the case of $^{240}$Pu 
the inclusion of pairing without APNP(LN) has only a marginal effect on the 
charge quadrupole and mass hexadecupole moments of inertia 
(Table 1). On the contrary, APNP(LN) has a strong 
impact on these moments decreasing their values. 
In addition, APNP(LN) increases the strength of pairing 
correlations especially for the neutron subsystem. In the CRHB 
theory, the pairing energies are defined as 
$E_{pairing}=-\frac{1}{2}Tr(\Delta \kappa)$ \cite{CRHB}. The 
values of neutron and proton pairing energies 
obtained in the CRHB calculations without and with 
APNP(LN) are $E_{pairing}^{\nu}=-1.823$ MeV, $E_{pairing}^{\pi}=-10.030$ MeV 
and $E_{pairing}^{\nu}=-13.129$ MeV, 
$E_{pairing}^{\pi}=-13.603$ MeV, respectively.

\vspace*{-18pt}
\begin{table}[h] 
\caption[]{Experimental and theoretical charge quadrupole moments of 
SD fission isomers. The results of the calculations with the NL1 and 
NL3 forces are given. Experimental data for U and Pu isotopes are 
taken from Ref.\ \protect\cite{H.89}, while the one for $^{242}$Am from
Ref.\ \protect\cite{Am242}.}
\begin{center}
\begin{minipage}{1.0\textwidth} 
\renewcommand{\footnoterule}{\kern -3pt} 
\begin{tabular}{ccccccc}
\hline\\[-10pt] 		

                   & $^{236}$U & $^{238}$U & $^{236}$Pu & $^{239}$Pu &
                   $^{240}$Pu & $^{242}$Am \\
\hline\\[-10pt]			
$Q_0^{exp}$ ($e$b) & $32\pm 5$ & $29\pm 3$ & $37\pm 10$ & $36\pm 4$ &
                   & $35.5\pm 1.0_{st}\pm 1.2_{mod}$ \\
$Q_0^{\rm NL1}$ ($e$b) &  35.9 & 37.0 & 36.6 & & 38.2 & \\
$Q_0^{\rm NL3}$ ($e$b) &  33.8 & 34.0 & 35.0 & & 35.5 & \\
\hline 
\end{tabular}
\end{minipage}
\renewcommand{\footnoterule}{\kern-3pt \hrule width .4\columnwidth 
\kern 2.6pt} 			
\end{center}
\end{table}
\vspace*{-8pt}

  The charge quadrupole moments calculated with the forces NL1 and NL3 
are compared with available experimental data in Table 2. One should 
note that the small error bars on the experimental values of $Q_0$ given for 
$^{238}$U and $^{240}$Am nuclei should be treated with caution since 
even modern experiments do not provide an accuracy of the absolute 
$Q_0$ values better than 15\%, see discussion in Ref.\ \cite{A190}. In 
addition, when comparing the calculations with experiment one should
take into account that {\bf (i)} the $Q_0^{exp}$ values have been
obtained with different experimental techniques \cite{H.89}, {\bf
(ii)} it is reasonable to expect that an addition of one neutron to 
$^{239}$Pu will not change considerably the $Q_0$ value and thus 
$Q_0^{exp}$($^{239}$Pu) could be used for comparison with the
calculated $Q_0$($^{240}$Pu). With these considerations in mind, it is 
clear that the results of the calculations come reasonably close to 
experiment.

  In conclusion, the cranked relativistic Hartree-Bogoliubov theory
has been applied for the description of the rotational and deformation
properties of superdeformed fission isomers in the $A\sim 240$ mass
region. This theory does not employ any adjustable parameters and
for the description of pairing correlations uses the well established
D1S Gogny force. The study of the dependence of the results on the
parametrization of the RMF Lagrangian reveals that rotational 
properties are best described by the NL1 force, while the
uncertainties of the experimental data on charge quadrupole
moments do not allow to use this experimental quantity for the
selection of the best RMF force.
  
 

   A.V.A. acknowledges support from the Alexander von Humboldt
Foundation. This work is also supported in part by the
Bundesministerium f{\"u}r Bildung und Forschung under the 
project 06 TM 979.
 
\begin{notes}
\item[a]
Alexander von Humboldt fellow, on leave of absence from the Laboratory 
of Radiation Physics, Institute of Solid State Physics, University of 
Latvia, LV 2169 Salaspils, Miera str.\ 31, Latvia
\end{notes}

\vfill\eject
\end{document}